\documentclass[aps,twocolumn,amsmath,amssymb,prb]{revtex4-2}
\usepackage{graphicx}
\usepackage{dcolumn}% Align table columns on decimal point
\usepackage{bm}% bold math
\usepackage{amsmath}

\usepackage{soul}
\usepackage{color}
\usepackage[colorlinks = true, allcolors = blue]{hyperref}

\begin{document}
\title{Four-terminal voltage fluctuations in disordered graphene nanoribbons:\\ Anderson and anomalous localization effects}
\author{Pablo Encarnaci\'on and  Victor A. Gopar}
\affiliation{$^1$Departamento de F\'isica Te\'orica and BIFI, Universidad de Zaragoza, Pedro Cerbuna 12, E-50009, Zaragoza, Spain}
%\date{.}
\begin{abstract}
Voltage is  a sensitive quantity to quantum interference in coherent electronic transport. We study the voltage fluctuations in disordered graphene nanoribbons with zigzag and armchair edge terminations in a four-terminal configuration. We show that
the average and standard deviation of the voltage oscillates with the separation of the attached voltage probes and depend on the coupling strength of the probes. The voltage fluctuations can be
large enough to observe negative voltages for weakly coupled probes. As we numerically verified, the voltage fluctuations are described within a random matrix approach for weakly disordered nanoribbons at energies away from the Fermi energy. However, near the Fermi energy, zigzag nanoribbons exhibit Anderson localization, whereas  electrons are anomalously localized in armchair nanoribbons. This distinction leads to different voltage statistics for zigzag and armchair nanoribbons.
\end{abstract}
%\pacs{}
\maketitle
\section{Introduction}
Quantum interference effects on the electronic transport at the mesoscopic and nanoscopic scales have been extensively  investigated for decades since the first fabricated electronic devices in which the wave nature of electrons was manifested in  quantities such as the conductance and the voltage~\cite{Washburn}.

Two-terminal configurations, where samples have a lead connected to electron reservoirs with fixed electrochemical potentials on each side have been helpful in understanding several interference effects on quantum electronic transport.   
For instance, it was  recognized that interference processes in ordinary disordered metals lead to the exponential localization in real space of the wave function of electrons, namely, the Anderson localization~\cite{Anderson58}, and that the complexity of the scattering processes in disorder leads to universal transport properties  such as the celebrated universal conductance
fluctuations~\cite{Umbach1984,Lee1985,Altshuler1985,Licini1985}.

Four-terminal configurations, however, are standard setups for measuring transport quantities, such as the resistance of materials. One of the advantages of these configurations over the simpler two-terminal setups is that the contact resistance is avoided. 
Additionally, four-terminal configurations  can be used to measure the voltage drop by adjusting the electrochemical potentials of two of the terminals acting as voltage probes until currents in these terminals vanish \cite{Buttiker1985,Tarucha,Buttiker1988}. Four-terminal configurations  have shown  surprising nonlocal effects on the voltage since pioneering quantum transport experiments~\cite{Takagaki,Picciotto}, and,  similarly to the conductance, voltage fluctuations in disordered media have been studied since the 1980s~\cite{Washburn,Lee1985,Benoit1987,Maekawa1987,Skoqpol1987,Roukes1987,Kane1988, Feng1991,Zhang2006}. 

Recent experimental progress in the fabrication of mesoscopic and nanoscopic circuits and scanning tunneling microscopy advances  have opened the possibility of spatially resolve electronic transport at the nanoscale by performing 
transport measurements between probes with a few nanometers of separation in four-terminal configurations. See, for instance,  Refs.~\cite{Nakayama,Ping2013} for a review.

The electronic transport in graphene-based structures has been under intensive research since the first production of graphene samples by exfoliation of graphite~\cite{Novoselov2004,Novoselov2005}. The electronic properties of graphene near the Fermi energy have been particularly interesting  
from practical and fundamental points of view because  electrons behave like relativistic particles with a linear dispersion relation.  
Nowadays, the peculiar properties of the honeycomb lattice of graphene 
have inspired investigations beyond electron wave transport, including investigations of microwave   transport and flexural edge states in graphene-like structures~\cite{Matthieu2013,Huang2024}.

As previously mentioned, four-terminal setups are commonly used to 
investigate  electronic transport in different materials, and graphene is no   exception.  For example, graphene nanoribbons  
can be contacted with probes made of normal metals, single-carbon nanotubes, or graphene~\cite{Ryo2012, Chen2015,Shi2015,Zhang2023}. 
Also, prior to the first fabrication of graphene samples, the resistance and voltage of carbon nanotubes were experimentally studied. Among the different quantum interference effects on the resistance and the voltage investigated, it was shown that negative resistances can be measured in carbon nanotubes in four-terminal configurations with noninvasive voltage probes~\cite{Gao2005, Gunnarsson, Makarovski, Gao2005}.

It is known that the termination graphene-nanoribbon's edges play a crucial role in their electronic properties. Additionally, graphene-based devices show particular transport properties in the presence of sources of disorder, such as lattice deformations or impurities in the lattices. For instance, in disordered armchair nanoribbons near the Fermi energy, electrons are weakly localized compared to the exponential localization seen in the widely known Anderson localization phenomenon. In contrast, electronic edge states in zigzag nanoribbons are exponentially localized in the presence of disorder. This is an example of the crucial role of nanoribbons' edges on the electronic properties. See Refs.~\cite{Peres,Sarma,Cooper2011,Foa_book} and
references therein for a review of the topic.

In this paper, we studied the random fluctuations of the voltage in disordered nanoribbons with zigzag and armchair edge terminations in a four-terminal setup. Two of the four terminals acting as voltage probes are attached to the ends of the nanoribbons, as shown in Fig.~\ref{fig_1}. Thus,  the voltage fluctuations are investigated as a function of the distance between the voltage probes and the strength of the probes' coupling to the nanoribbons using the scattering approach to quantum transport and random-matrix theory~\cite{Buttiker1985,Mello1992, Beenakker_RMP,Mello-book,Texier_2016}. 

Random-matrix models have  described   several statistical transport properties  of disordered conductors in the presence of Anderson localization~\cite{Beenakker_RMP,Mello-book}. For example, 
it has been found that the average logarithmic conductance is a linear function of the system length:$\langle -\ln G \rangle=L/\ell$ (the conductance $G$ is in units of the conductance quantum $2e^2/h$).
In contrast,  a power-law dependence of the average logarithmic conductance with the length $L$: $\langle -\ln G \rangle \sim L^{\alpha}$ was found in armchair nanoribbons 
due to presence of anomalous localization, near the Fermi energy~\cite{Ioannis2013}. As shown below, standard Anderson and anomalous localizations have distinct and strong effects on voltage fluctuations.

We perform numerical simulations using the tight-binding model to calculate the voltage and conductance of metallic armchairs and zigzag nanoribbons away and near the Fermi energy. 
The statistical properties of the voltage obtained from the numerical results are contrasted with those of random matrix theory (RMT). Additionally, we  conduct numerical simulations of one-dimensional (1D) disorder systems to obtain the voltage statistics and show that the effects of the disorder on the voltage fluctuations presented here are not restricted to graphene nanoribbons but, as we show below, the voltage statistics are essentially consequences of the general phenomenon of  electron localization.

\section{Scattering approach to the voltage in a four-terminal setup}
\label{section_2}

\begin{figure}
\centering
\includegraphics[width=\columnwidth]{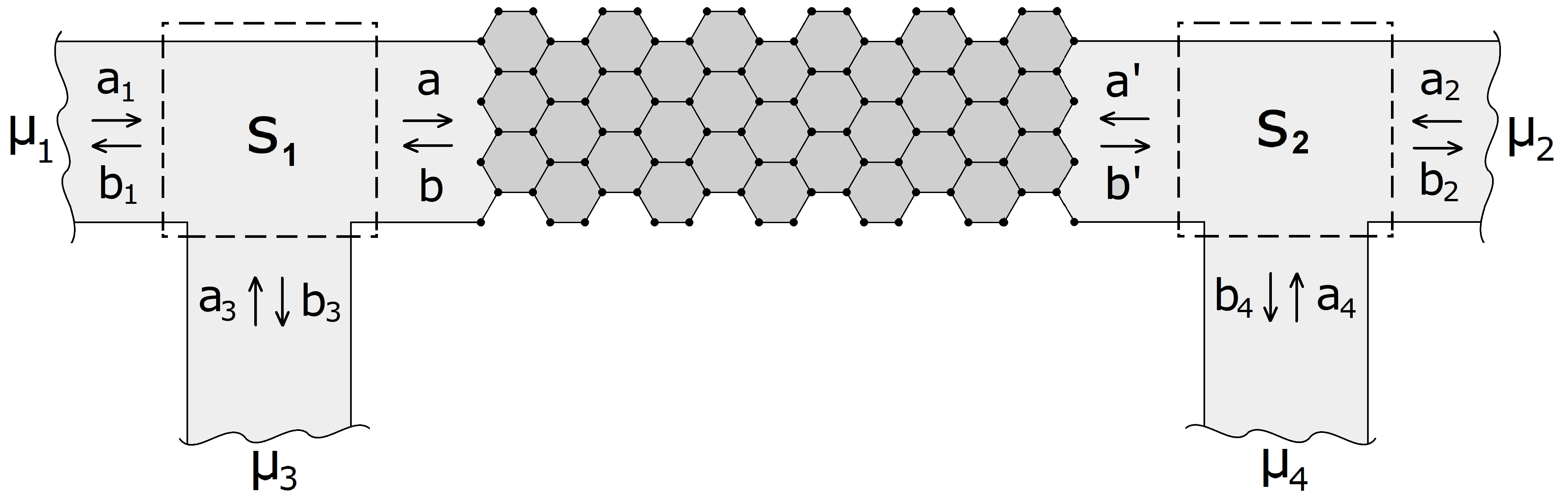}
\caption{Schematic of a four-terminal configuration. A current flows between the perfect leads attached along the entire width of the nanoribbon and 
to electron reservoirs with electrochemical potentials $\mu_1$ and $\mu_2$, while two terminal acting as voltage probes are attached to reservoirs with electrochemical potentials $\mu_3$ and $\mu_4$. The scattering matrix $S_G$ relates the amplitudes ($a,b$) and ($a',b'$) at each side of the nanoribbon as given in Eq.~(\ref{SG}).}
\label{fig_1}
\end{figure}

Let us consider a nanoribbon with  four terminals attached, as shown in Fig.~\ref{fig_1}. A current flow from terminals 1 and 2, while terminals 3 and 4 are used as voltage probes. 
Within the Landauer-B\"uttiker approach to electronic transport, it is assumed that 
a current $I$ flows between the contacts on the left and right side of the sample, which are at electrochemical potentials $\mu_1$ and $\mu_2$, respectively. The chemical potentials, 
or voltages $V_i=\mu_i/e$, where $e$ is the electron charge, in terminals 3 and 4 are fixed by the condition that no current flows in these two terminals. Thus, the voltage $V$ between the terminals 3 and 4 is defined by 
the ratio of the difference of the electrochemical potentials: $V={\left(\mu_3-\mu_4 \right)}/{\left( \mu_1-\mu_2 \right)}$, and is given in terms of the transmission between the terminals as~\cite{Buttiker1985} 
\begin{equation}
\label{voltage}
 V= \frac{T_{31}T_{42}-T_{32} T_{41}}{\left(T_{31}+T_{32}+T_{34} \right)\left(T_{41}+T_{42}+T_{43}  \right)-T_{34}T_{43}},
\end{equation}
where $T_{ij}$ is the transmission coefficient between the terminals $j$ and $i$. 

We now introduce a model of the scattering matrix $S$ of the four-terminal setup, see  Fig.~\ref{fig_1}, whose elements provide the  transmission coefficients $T_{ij}$ that we need to calculate the voltage, according to  Eq.~(\ref{voltage}).  
We first assume that the incoming ($a,a'$) and outgoing ($b,b'$) wave amplitudes at both sides of the graphene nanoribbon are related by the scattering matrix:  
\begin{equation}
\label{SG}
S_G= \left( \begin{matrix}
  r & t' \\
  t & r'
 \end{matrix}\right),
\end{equation}
where $r$ and $r'$ are the reflection amplitudes, while $t$ and $t'$ are the transmission amplitudes of incoming and outgoing waves to the nanoribbon. We will assume time-reversal symmetry in the following, thus $t=t'$. However, it is convenient to express the scattering matrix $S_G$ in terms of the transfer matrix $M_G$, that relates the wave amplitudes on the left ($a,b$) and on right side ($a',b'$) of the nanoribbon: 
\begin{equation}
\label{MG}
\left( \begin{matrix}
  a' \\
  b' \\
 \end{matrix}\right)=M_G
 \left(\begin{matrix}
  a \\
  b \\
 \end{matrix}\right) \ \ \mathrm{with} \ \ 
M_G= \left( \begin{matrix}
  \alpha & \beta \\
  \beta^* & \alpha^*
 \end{matrix}\right),
\end{equation}
where $\alpha$ and $\beta$ are complex numbers that satisfy $|\alpha|^ 2- |\beta|^2=1$. In terms of $M_G$, the elements of $S_G$ in Eq~(\ref{SG}) are given by $r=-\beta^*/\alpha^*$, $t=1/\alpha^*$, and $r'=\beta/\alpha^*$.

We note that $S_G$, and $M_G$, as introduced in Eqs.~(\ref{SG}) and (\ref{MG}), are $2\times 2$ matrices. We made this assumption because we are interested in energies in the vicinity of the Fermi energy for both metallic armchair and zigzag graphene nanoribbons where a single channel contributes to the transport. 

For the scattering matrices of the nodes $S_1$ and $S_2$  
 (see Fig.~\ref{fig_1}), we introduce a model~\cite{Buttiker1985} that depends on the parameter 
 $\varepsilon$, which is interpreted as the strength coupling of the voltage probes  attached to the nanoribbon 
\begin{equation}
S_{1}=S_{2}= \left( \begin{matrix}
  a & b & \sqrt{\varepsilon} \\
  b & a & \sqrt{\varepsilon} \\
  \sqrt{\varepsilon} & \sqrt{\varepsilon} & -\sqrt{1-2\varepsilon}
 \end{matrix}\right),
\end{equation}
where $a$, $b$, and $\varepsilon$ are real numbers with $0 \le \varepsilon \le 1/2$. From the unitary property of the scattering matrices $S_{1}$ and $S_{2}$, the elements $a$ and $b$ are given by $a=-\left(1-\sqrt{1-2\varepsilon}\right)/2$, $b=\left(1+\sqrt{1-2\varepsilon}\right)/2$. 

Finally, the scattering matrix $S$ of the complete four-terminal configuration  that relates incoming and outgoing waves of the four-terminal configuration, i.e.,  
\begin{equation}
\left( \begin{matrix}
  b_1 \\
  b_2 \\
  b_3 \\
  b_4
 \end{matrix}\right)=S \left(\begin{matrix}
  a_1 \\
  a_2 \\
  a_3 \\
  a_4
 \end{matrix}\right),
\end{equation}
is given by the combination of the scattering matrices of the nodes and the nanoribbon: $S=S_1\otimes S_G\otimes S_2$. The transmission coefficients $T_{ij}$ in  Eq~(\ref{voltage}) are  given by the elements $s_{ij}$ of the $S$-matrix: $T_{ij}=|s_{ij}|^2$.

Thus, combining the scattering matrices $S_G$, $S_{1}$, and $S_{2}$, the transmission coefficients are given by~\cite{Godoy,Gopar1994}
\begin{equation}
\label{Sij}
T_{ij}=\frac{\varepsilon}{|\alpha|^4|\Delta|^2}|c_{ij}|^2 ,  
\end{equation}
where $\Delta=\left(r^*-a \right)\left(r'^*-a \right)-{t^{*}}^{2}$, and according to Eq.~(\ref{voltage}), we need the following coefficients:  
\begin{eqnarray}
\label{Cij}
 c_{31}&=& \left( a \beta + \beta^{*}\right)\alpha e^{-ik_F L}- |\alpha|^2 e^{-2ik_F L}-a\alpha^2 , \nonumber \\
 c_{42}&=&\left( \beta + a\beta^{*}\right)\alpha e^{-ik_F L}+ |\alpha|^2 e^{-2ik_F L}+a\alpha^2, \nonumber  \\
 c_{41}&=&c_{32}=b|\alpha| , \hspace{0.4cm} c_{43}=\sqrt{\varepsilon} |\alpha| ,
\end{eqnarray} 
where $k$ is the wave vector and $L$ is the distance between the voltage probes.
We can verify that for a perfect conducting sample ($\alpha=1$, $\beta=0$) with vanishing coupling $\varepsilon=0$,  there is no voltage drop, $V_{\alpha=1,\beta=0}=0$. 
However,  for nonzero coupling, the voltage oscillates with the distance between the voltage probes: using Eqs.~(\ref{Sij}) and  (\ref{Cij}) with $\alpha=1$ and $\beta=0$, we have
\begin{eqnarray}
\label{Tijperfect}
 T_{31}&=&T_{42}=\frac{\varepsilon}{\left|\Delta \right|^2}\left|a+e^{-2ik_FL} \right|^2 , \hspace{0.4cm} T_{32} = T_{41}=\frac{\varepsilon}{\left|\Delta \right|^2}b^2, \nonumber  \\ 
 T_{34}&=&T_{43}=\frac{\varepsilon}{\left|\Delta \right|^2}\varepsilon. 
\end{eqnarray}
Substituting these expressions, Eq.~(\ref{Tijperfect}), in Eq.~(\ref{voltage}), we obtain for a perfect conductor that 
\begin{eqnarray}
\label{voltage_perfect}
 V_{\alpha=1,\beta=0}&=&\frac{\left|a+e^{-2ik_F L}\right|^2-b^2}{\left|a+e^{-2ik_F L}\right|^2+b^2+2\varepsilon} \nonumber \\
&=&\frac{\left(1-\sqrt{1-2\varepsilon}\right)\left(1- \cos(2k_F L)  \right)}{2+\varepsilon-\left(1 - \sqrt{1-2\varepsilon}\right)\cos(2k_F L)} . 
\end{eqnarray} 
Note that the oscillatory behavior of the voltage with the distance between the voltage probes in Eq.~(\ref{voltage_perfect}) illustrates an interference effect in the voltage in the four-terminal setup. For future reference, we define $V_{0,\mathrm{max}}$ and $V_{0,\mathrm{min}}$ as the  maximum and  minimum, respectively, of the oscillations of $V_{\alpha=1,\beta=0}$ as described by  Eq.~(\ref{voltage_perfect}), at their respective $k_FL$ values.

\section{RMT analysis of the average and variance of voltage}
\label{RMT_section}

So far, we have not considered any source of disorder, such as impurities or distortions in the lattice structure of the nanoribbons. The presence of disorder leads to complex scattering processes of electrons with observable effects at the level of transport quantities, such as random fluctuations of the conductance and the voltage. 
From  Eq.~(\ref{voltage}), we notice that    
the voltage fluctuations will be determined by the statistics of the transmission coefficients between the different terminals when disorder is introduced in  the nanoribbons.

Several electronic transport properties in disorder media  have been successfully studied within RMT, although most of these properties have been investigated in two-terminal setups. See Refs. ~\cite{Beenakker_RMP, Mello-book} for a review of the topic. Statistical properties of the voltage in three and four-terminal setups have  been studied within the so-called maximum entropy approach \cite{Godoy,Gopar1994}; see also Ref.~\cite{Texier_2016}. Within this approach, it has been derived an evolution equation for the expectation of functions of elements of the transfer matrices 
$ \langle f(\alpha, \alpha^*,\beta,\beta^*) \rangle$ with the system length $L$ in the so-called weak scattering limit. For 1D systems, the evolution differential equation is given by~\cite{Mello1992,Mello-book}:
\begin{eqnarray}
\label{diffusion_M}
  && \frac{\partial \langle f \rangle}{\partial s}   \nonumber \\  
  && = \bigg{\langle} \left\{ (\alpha\alpha^* + \beta\beta^*)\left[ \frac{\partial^2}{\partial\alpha\partial\alpha^*} +\frac{\partial^2}{\partial\beta\partial\beta^*}\right] +  \right.
    \left\{ 2\alpha\beta^*\frac{\partial^2}{\partial\alpha\partial\beta^*} \right. \nonumber  \\ 
    & - & \left. \left. \frac{1}{2} \left[ \alpha^2\frac{\partial^2}{\partial\alpha^2} + 2\alpha\beta \frac{\partial^2}{\partial\alpha\partial\beta} + \beta^2\frac{\partial^2}{\partial\beta^2} \right] \right\} +\{\mathrm{c.c.} \} \right\} f\bigg{\rangle}, \nonumber \\ 
\end{eqnarray} 
where $s=L/\ell$, and $\ell$ is the mean free path. $\{\mathrm{c.c.}\}$ stands for the complex conjugate of the terms in the inner  curly brackets.  Notice that 
the voltage in Eq.~(\ref{voltage}) can be written as a function of the elements
$\alpha, \alpha^*,\beta, \beta^*$ using Eqs.~(\ref{voltage}), (\ref{Sij}), and (\ref{Cij}). Thus, we can set $f(\alpha, \alpha^*,\beta,\beta^*)=V(\alpha, \alpha^*,\beta,\beta^*)$ to obtain $\langle V \rangle$ or $f(\alpha, \alpha^*,\beta,\beta^*)=V^2(\alpha, \alpha^*,\beta,\beta^*)$ and solve Eq.~(\ref{diffusion_M}) to calculate the second moment $\langle V^2\rangle$. 
One of the main results of this approach is that all the statistical properties of functions of the transfer matrices are determined by the single parameter $s$.

\begin{figure*}
\includegraphics[width=1.9\columnwidth]{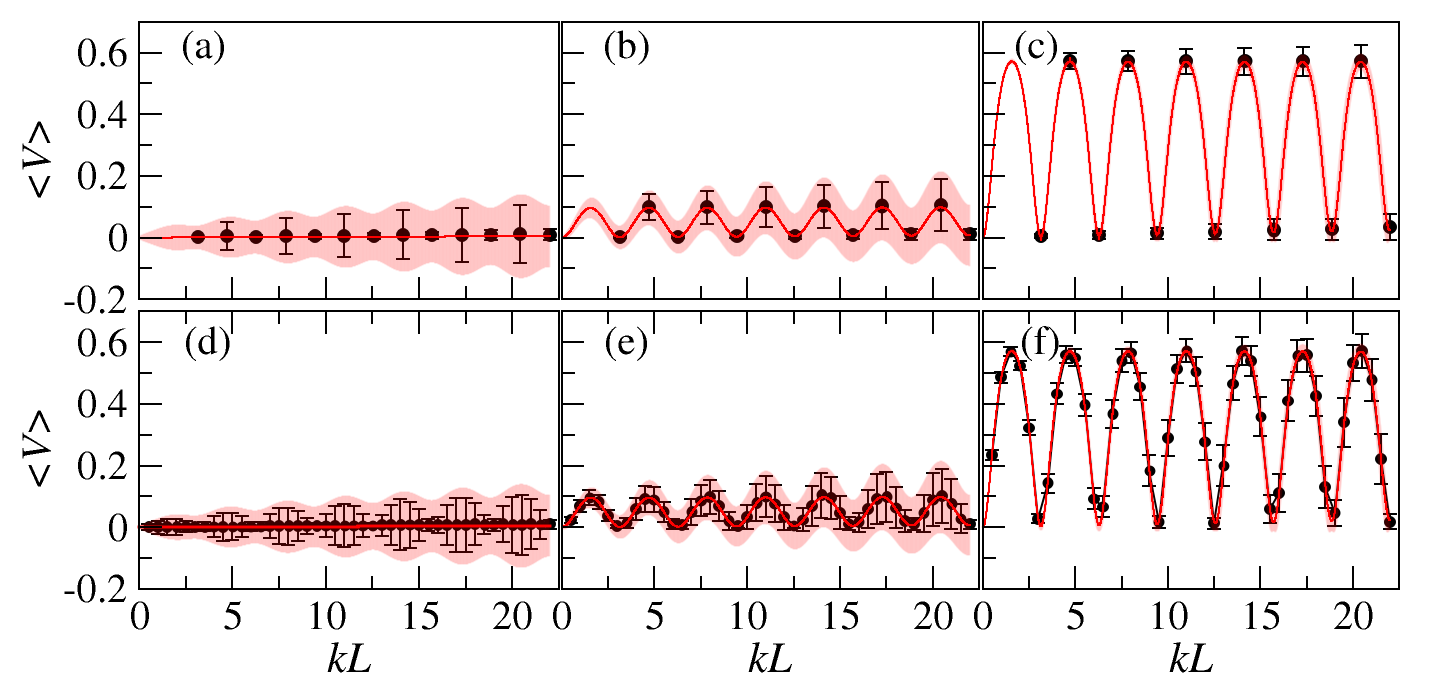}
\caption{Theoretical (red lines and shaded regions) and numerical (black dots and error bars) results of $\langle V \rangle$ and $\delta V$, for $\epsilon = 0$ (a, d), $0.1$ (b, e) and $0.5$ (c, f). Top panels (a-c) show the results for zigzag nanoribbons for $ka=\pi/2$, $l/a=719.5$ and $W=0.17$. Bottom panels (d-f) show the results for armchair nanoribbons for  $ka=0.289$, $l/a=3698.0$ and $W=0.09$. Numerical results are obtained averaging over 5000 samples.}
\label{fig_2}
\end{figure*}

The general expression of $V(\alpha, \alpha^*,\beta,\beta^*)$ as given by Eq.~(\ref{voltage}) is very involved and an analytical solution of Eq.~(\ref{diffusion_M}) for arbitrary degree of disorder and coupling $\varepsilon$ is  not possible. However, in the ballistic limit, $L \ll \ell$, 
Eq.~(\ref{diffusion_M}) is simplified by expanding $V(\alpha,\alpha^*,\beta,\beta^*)$ around a perfect conductor: $\alpha=1, \beta=0$.  In this case, it is found that the average voltage $\langle V \rangle$ and the variance voltage  $\mathrm{Var}(V)= \langle (V-\langle V\rangle)^2 \rangle $ are given by, up to  the first order in $L/\ell$~\cite{Gopar1994}: 
\begin{eqnarray}
\label{averageV}
 \langle V  \rangle& = & V_{\alpha=1,\beta=0} + \frac{L}{\ell} \left[\frac{\partial^2 V}{\partial \alpha \partial \alpha^*} + \frac{\partial^2 V}{\partial \beta \partial \beta^*} -\frac{1}{2}\frac{\partial^2 V}{\partial\alpha^2} \right. \nonumber \\
 &-& \left. \frac{1}{2}\frac{\partial^2 V}{\partial{\alpha^*}^2} \right]_{\alpha=1,\beta=0}
\end{eqnarray}
and 
\begin{eqnarray}
\label{varV} 
 \mathrm{Var}(V) &= &\frac{L}{\ell}  \left[2\left(\frac{\partial V}{\partial \alpha}\frac{\partial V}{\partial \alpha^*}
 + \frac{\partial V}{\partial \beta} \frac{\partial V}{\partial \beta^*}\right) \right. \nonumber \\  
& -&\left. \left(\frac{\partial V}{\partial\alpha}\right)^2-\left(\frac{\partial V}{\partial{\alpha^*}}\right)^2 \right]_{\alpha=1,\beta=0},
\end{eqnarray}
respectively. For vanishing coupling, the above expressions, Eqs.~(\ref{averageV}) and (\ref{varV}), are further reduced to:
\begin{eqnarray}
\label{averageVnull}
 \langle V \rangle & = &\frac{1}{2} \frac{L}{\ell}, \\ \label{varianceVnull}
 \mathrm{Var}(V)& = &\frac{1}{4} \frac{L}{\ell}\sin^2\left(k L \right). 
\end{eqnarray}
Thus, the average voltage is a linear function of the probe separation, while the variance oscillates with the separation. Also, we notice that the standard deviation $\delta V=\sqrt{\mathrm{Var}(V)}$ can be larger than $\langle V \rangle$,  which opens the possibility of observing negative voltages. 
For arbitrary coupling strength, we can use the expressions in Eqs.~(\ref{averageV}) and (\ref{varV}). 

In the next section, we show the results from the numerical evaluation of the expressions for $\langle V \rangle$ and the variance $\mathrm{Var}(V)$, as given by  Eqs.~(\ref{averageV})  and  (\ref{varV}), respectively, for finite values of the coupling $\varepsilon$. We thus compare these theoretical results with those from numerical simulations of graphene nanoribbons.

\section{Voltage fluctuations in armchair and zigzag nanoribbons}

We use the standard tight binding model to obtain the scattering matrix $S_G$ or, equivalently, the transfer matrix $M_G$. The single-particle tight-binding Hamiltonian is given by 
\begin{equation}
 H=\sum_{\langle i,j\rangle}\tau_{ij}\left(n_i^\dagger n_j +n_j^\dagger n_i  \right),
\end{equation}
where the sum is over nearest neighbors $i$, $j$, and $n_i^\dagger$ ($n_i$) is the creation (annihilation) operator for spinless fermions and $\tau_{ij}$ is the hopping that connects the two graphene sublattices. For pristine graphene $\tau_0=2.74$eV. The numerical simulations were performed using the Kwant code~\cite{kwant} 
for metallic nanoribbons. We recall that the width of the nanoribbons determines the band structure~\cite{Wakabayashi2009}.  We have set the width of zigzag nanoribbons to $ 13 a/\sqrt{12}$, which corresponds to 5 zigzag chains, while for metallic armchair nanoribbons, the width is fixed to $7a/2$, where $a=2.46$\r{A} is the graphene lattice constant. 
The disorder is introduced via random hopping elements $\tau_{i,j}$ sampled from the uniform distribution in the interval $[\tau_0-W,\tau_0+W]$. Thus, by changing the strength of the disorder $W$, we can control the transport regime of the nanoribbons. This type of disorder, namely off-diagonal disorder, preserves the symmetry of the graphene lattice and can be interpreted as a model describing the presence of random distortions 
in the graphene lattice.  
Different realizations of the disorder are generated to obtain the average $\langle V \rangle$ and the variance $\mathrm{Var}(V)$, as well as the complete distribution of the voltage $P(V)$.

We divide the analysis of the voltage fluctuations in two different scenarios; those that occur near  the Fermi energy and those that occur far from it. As we mentioned in the introductory part, graphene nanoribbons behave like conventional conductors away from the Fermi energy, whereas  near the Fermi energy, graphene nanoribbons exhibit their peculiar electronic properties~\cite{Wakabayashi2009}. In both energy regimes, we restrict to energies where a single channel contributes to the transport.

\begin{figure*}[tpb]
\includegraphics[width=2\columnwidth]{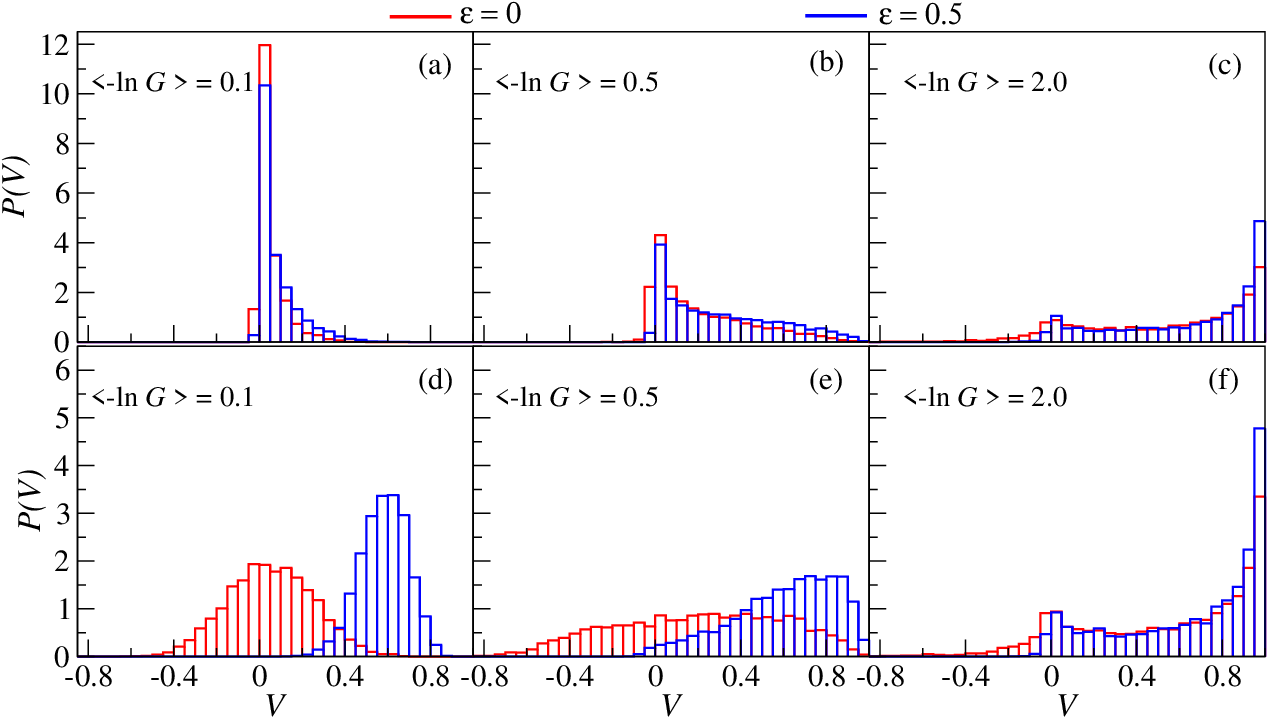}
\caption{Voltage distribution $P(V)$ for zigzag nanoribbons at the Fermi energy for three different values of $\langle -\ln G \rangle$ and two different values of the coupling: $\varepsilon=0$ (red histograms) and $\varepsilon=0.5$ (blue histograms). Top panels are obtained for a fixed $ka=3.09$ and $kL=74.2$, where the voltage reaches its minimum value in absence of disorder, $V_{0,\mathrm{min}}$, and $W=0.223, 0.504, 1.021$ for panels (a), (b), and (c), respectively. Bottom panels are obtained for a fixed $ka=3.09$ and $kL=358.4$, where the voltage reaches its maximum value in absence of disorder, $V_{0,\mathrm{max}}$, and $W=0.504, 1.179, 2.236$ for panels (d), (e), and (f), respectively. All distributions are  obtained from 5000 disorder realizations.}
\label{fig_3}
\end{figure*}

\subsection{Voltage fluctuations away from the Fermi energy}

To compare with the previous results of RMT in Sec.~\ref{RMT_section}, numerical simulations are conducted in the ballistic transport regime. The voltage probes are connected to the disordered nanoribbons and are separated by a distance $L$ following the model in Sec.~\ref{section_2} (see Fig.~\ref{fig_1}).

In Fig.~\ref{fig_2}, we show the results of the average 
$\langle V \rangle$ and the standard deviation $\delta V=\sqrt{\mathrm{Var}(V)}$  for zigzag (upper panels) and armchair (lower panels) nanoribbons for three different values of the voltage-probes coupling $\varepsilon$. The theoretical predictions for  $\langle V \rangle$ and $\delta V$, Eqs.~(\ref{averageV}) and (\ref{varV}),  are represented by red lines and shaded regions, respectively, while the numerical results are shown in black dots and error bars.
Notice that we use the dimensionless quantity $kL$ on the $x$ axis for convenience.  

Specifically, for negligible coupling strength $\varepsilon$,  as illustrated in Figs.~\ref{fig_2}(a) and \ref{fig_2}(d), 
$\langle V \rangle$ is a linear function of $L$, consistent with Eq.~(\ref{averageVnull}), and the standard deviation oscillates according to Eq.~(\ref{varianceVnull}). When there is a finite 
coupling of the voltage probes to the nanoribbons, both $\langle V \rangle$ and $\delta V$ oscillates  with the distance separation of the probes and the size of the fluctuations $\delta V$ are reduced with coupling $\varepsilon$. We also noted that the minimum values of $\langle V \rangle$ are practically unaffected by the coupling $\varepsilon$, in contrast to the maximums of $\langle V \rangle$, which increase as the coupling value is raised.

Overall, we can observe from Fig.~\ref{fig_2} that the RMT model captures  the trend of the average and variance voltage, although the voltage standard deviation is  slightly overestimated by the theoretical model for small values $\varepsilon$, whereas for the maximum 
coupling $\varepsilon=1/2$, $\delta V$ is underestimated.

We remark that the amount of disorder in zigzag and armchair nanoribbons has been chosen to ensure that they have the same mean free path $\ell$ ($k\ell \approx 1100$). Consequently,  both nanoribbons have the same ratio $ L/\ell$ for a given separation $L$,. In this manner, as shown in Fig.~\ref{fig_2}, the results for zigzag and armchair nanoribbons coincide. This reveals that the voltage statistics are determined solely by the ratio $L/\ell$, in agreement with RMT.

\subsection{Voltage fluctuations near the Fermi energy}

Graphene nanoribbons show their distinct properties  at low energies. For instance, in pristine zigzag nanoribbons, the electron wave function is located along the edges of the nanoribbons and becomes exponentially localized in the presence of disorder. However, in armchair nanoribbons, there are no electronic  edge states but  electrons become  weakly  or anomalous  localized in the presence of disorder,  rather than exhibiting the exponential decay typical of  Anderson localization~\cite{Ioannis2013}. Thus, graphene nanoribbons allow studying the effects of both types of localizations on voltage fluctuations.
 
 In this section, we obtain numerically the complete distribution of the random fluctuations of the voltage.  At 
low energies, $k$ can be very small. This means that simulating long nanoribbons would be necessary to observe the structure of the voltage oscillations, as those shown in Fig.~\ref{fig_2}, especially for armchair nanoribbons. Thus, instead of that,  we obtain the distribution of $V$ for different values of $\langle \ln G \rangle$, at two different $kL$ values. 
As we have seen previously, $V$ oscillates as a function of the distance probes, see Eq.~(\ref{voltage_perfect}) and Fig~\ref{fig_2}. Therefore, we have chosen  to calculate the voltage distribution at the $kL$ values where the voltage reaches a 
maximum, $V_{0,\mathrm{max}}$, and a minimum, $V_{0,\mathrm{min}}$, in disorder-free zigzag and armchair nanoribbons. Also, in the presence of  Anderson localization, $\langle -\ln G \rangle =L/\ell$ and therefore, it determines the transport regime:  In the ballistic regime  $\langle -\ln G \rangle <1$, whereas in the insulating regime, $\langle -\ln G \rangle > 1$. Thus, we use $\langle -\ln G \rangle$ as a parameter to indicate the amount of disorder.

\begin{figure*}[tpb]
\includegraphics[width=2\columnwidth]{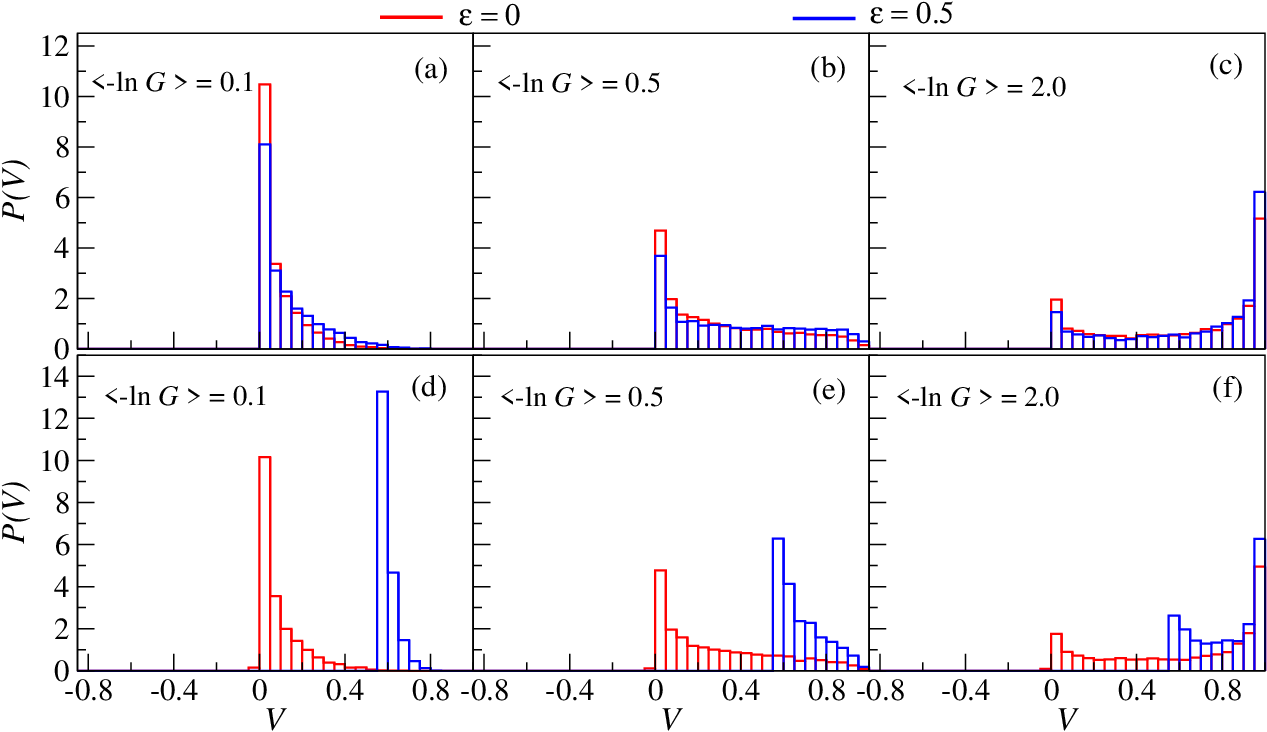}
\caption{Voltage distribution $P(V)$ for armchair nanoribbons at the Fermi energy for three different values of $\langle -\ln G \rangle$ and two different values of the coupling: $\epsilon=0$ (red histograms) and $\epsilon=0.5$ (blue histograms). Top panels are obtained for a fixed $ka=1.02\times  10^{-5}$, $kL=2.55\times 10^{-4}$, and $W=0.268,0.628,1.471$ for (a), (b), and (c), respectively, where the voltage reaches its minimum value in absence of disorder. Bottom panels are obtained for a fixed $ka=7.19\times 10^{-6}$, $kL=1.79\times 10^{-4}$, and $W=0.268,0.628,1.47$ for (d), (e), and (f), respectively, where the voltage reaches its maximum value in absence of disorder. All the distributions are obtained by generating 5000 different realizations of the disorder.}
\label{fig_4}
\end{figure*}

\subsubsection{Zigzag nanoribbons}

In Fig.~\ref{fig_3}, we show how the voltage distribution $P(V)$ changes for two  values of the coupling of the voltage probes: $\varepsilon=0$ (red histograms) and 1/2 (blue histograms), and three different values of $\langle \ln G \rangle$. 
The upper panels [Figs.~\ref{fig_3}(a), \ref{fig_3}(b), and \ref{fig_3}(c)], display the results for $P(V)$ calculated at a $kL$ value where the voltage reaches a minimum  in a pristine zigzag nanoribbon, $V_{0,\mathrm{min}}$. 
Conversely, in the lower panels [Figs.~\ref{fig_3}(d), \ref{fig_3}(e), and \ref{fig_3}(f)], show $P(V)$  obtained at the value of $kL$ at which the voltage reaches a maximum in a disorder-free zigzag nanoribbon. 

Various observations can be made regarding the results in Fig.~\ref{fig_3}. First, it is  seen that  the landscape of $P(V)$ for a fixed value of $\langle \ln G \rangle$ in the upper panels is  much less affected  by the strength coupling $\varepsilon$ than in the lower panels, especially when the disorder is not too strong ($\langle -\ln G \rangle <2$). 
For example, we can observe that the peak of the voltage distributions shifts to higher values of $V$, and the width of the distributions decreases as the strength of voltage probes coupling is increased Figs.~\ref{fig_3}(d) and \ref{fig_3}(e). This effect of the coupling strength has also been seen at energies far from Fermi energy: as we mentioned earlier, the results in Fig.~\ref{fig_2} show that the average voltage values at the minimums in the oscillatory behavior remain small when increasing the coupling strength, whereas the maximum values of the average voltage increase with the coupling strength and the deviation standard decreases, thereby reducing the probability of observing negative voltages.

Additionally, we observe that all the distributions coincide as the disorder amount increases even further; see Figs.~\ref{fig_3}(c) and ~\ref{fig_3}(f). This means that for strong  disorder,  electronic localization  overcomes the effects of the probe coupling.

\subsubsection{Armchair nanoribbons}

We now present the results for armchair nanoribbons. 
As we have pointed out previously, electrons are weakly localized in armchair nanoribbons  near the Fermi energy in relation to the exponential localization in the standard Anderson phenomenon. As shown below, anomalous localization also strongly affects the voltage fluctuations.

In Fig.~\ref{fig_4}, we present the voltage distributions for two different coupling probes $\varepsilon =0$ and $\varepsilon =1/2$ and three values of the disorder where $\langle \ln G \rangle$ take the values 0.1, 0.5, and 2.0. 
The distributions of the upper panels are calculated at a  $kL$ value at which the voltage reaches a minimum  in a pristine armchair nanoribbon. It is observed that the landscape of the voltage distributions remains largely unaffected  by the coupling of the voltage probes for a fixed value of $\langle \ln G \rangle$. 

The lower panels of Fig.~~\ref{fig_4} show the voltage distributions calculated at a $kL$ value at which the voltage reaches a maximum in disorder-free armchair nanoribbons. In contrast to the upper panels, it is noted that the voltage distributions are significantly affected by the coupling strength. As the coupling increases,  
the entire voltage distributions shift to higher values of voltages and become narrower. Consequently, the probability of negative voltages becomes zero, unlike the case of zigzag nanoribbons, as seen in the lower panels of Fig.~\ref{fig_3}.

When comparing the results in Figs.~\ref{fig_3} and \ref{fig_4}, we notice that the probability of negative voltages in armchair nanoribbons is negligible, in contrast to the case of zigzag nanoribbons with broader voltage distributions whose tails extend into negative voltage values.

The results above unveil that the statistics of the voltage fluctuations differ in zigzag and armchair nanoribbons near the Fermi energy, where standard Anderson and nonconventional localizations, respectively, take place. This suggests that the properties of voltage statistics are a  result of a more fundamental and general phenomenon, namely, wave localization, rather than particular properties of the graphene nanoribbons. To support this statement, we compared the voltage  distribution of 1D disordered systems in the presence of Anderson and anomalous localizations with the corresponding zigzag and armchair  nanoribbons results in the next section.

\section{Voltage fluctuations: Standard and anomalous localizations}

Anderson and anomalous localizations in 1D disordered wires can be studied 
using 1D tight-binding Hamiltonian models with off-diagonal disorder. In particular,  anomalous localization emerges near the center of the band~\cite{Theodorou1976,Antoniou1977,Soukoulis1981}. 
Thus, we perform simulations of 1D disordered wires in a four-terminal setup with two of the four attached leads separated by a distance $L$, which acts as  voltage probes, using the standard tight-binding Hamiltonian with random nearest-neighbor hopping. 

\begin{figure}
\includegraphics[width=\columnwidth]{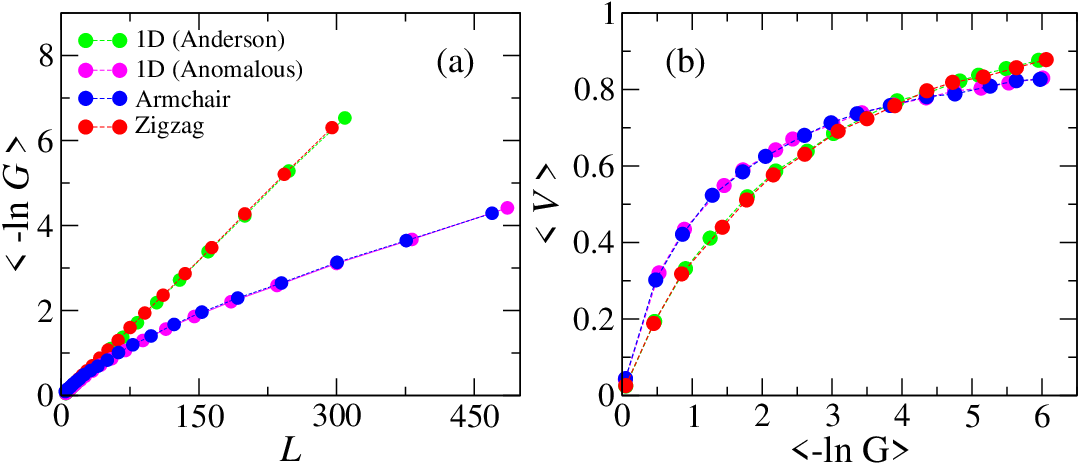}
\caption{(a) The average $\langle -\ln G \rangle$ is plotted against the system length $L$ for both Anderson and anomalous  localizations. For Anderson localization there is a linear dependence with $L$, while for anomalous localization, there is a power-law dependence, with power $\alpha=0.73$. (b) The average voltage plotted against $\langle -\ln G \rangle$ for $\varepsilon=0$ for zigzag and armchair nanoribbons near the Fermi energy, as well as for 1D wires in the presence of Anderson and anomalous localization. The disorder strength $W$ varies from 0.156 to 1.596 for zigzag nanoribbons and 0.094 to 1.678 for armchair nanoribbons.}
\label{fig_5}
\end{figure}

We first verify the presence of Anderson and anomalous localization in the 1D systems by obtaining the average of the logarithmic conductance as a function of the wire length. As expected, far from the band center, $\langle \ln G \rangle$ depends linearly with $L$, whereas near the band center $\langle \ln G \rangle$ follows a power law:  $\langle \ln G \rangle \propto L^\alpha$ 
with $\alpha=0.73$. Both dependencies are shown in Fig.~\ref{fig_5}(a). For the case of 1D wires, we have adjusted the energy to match the power $\alpha=0.73$ found previously for armchair nanoribbons. 
Additionally, we plotted $\langle \ln G \rangle$ for zigzag and armchair nanoribbons near the Fermi energy in Fig.~\ref{fig_5}(a). Our results indicate that 
Anderson and anomalous localizations are present in zigzag and armchair nanoribbons, and 1D  disordered wires. As shown in Fig.~\ref{fig_5}(a), the results for 1D wires and graphene nanoribbons coincide. 

Before presenting the results for the voltage distributions, a more straightforward quantity, the voltage average, can first signify the effects of both types of localizations in the voltage fluctuations.

In Fig.~\ref{fig_5}(b), we have plotted the average $\langle V \rangle$ against the degree of disorder, quantified by $\langle \ln G \rangle$, for 1D disordered wires in the presence of Anderson and anomalous localizations. In the same figure, $\langle V \rangle$ is plotted for zigzag and armchair nanoribbons near the Fermi energy.  
%The average voltage is calculated at values of $kL$ where the voltage is maximum.  
As shown, the average voltage of 1D wires in the presence of Anderson localizations matches $\langle V \rangle$ for zigzag nanoribbons. Similarly, the results of $\langle V \rangle$ for 1D wires  with anomalous localizations agree with the results for armchair nanoribbons. Additionally, from Fig.~\ref{fig_5}(b), it is seen that in the presence of Anderson localization, the average voltage approaches its limit value $\langle V \rangle=1$ more rapidly than in the presence of anomalous localization.

\begin{figure*}[tpb]
%\begin{center}
\includegraphics[width=2.1\columnwidth]{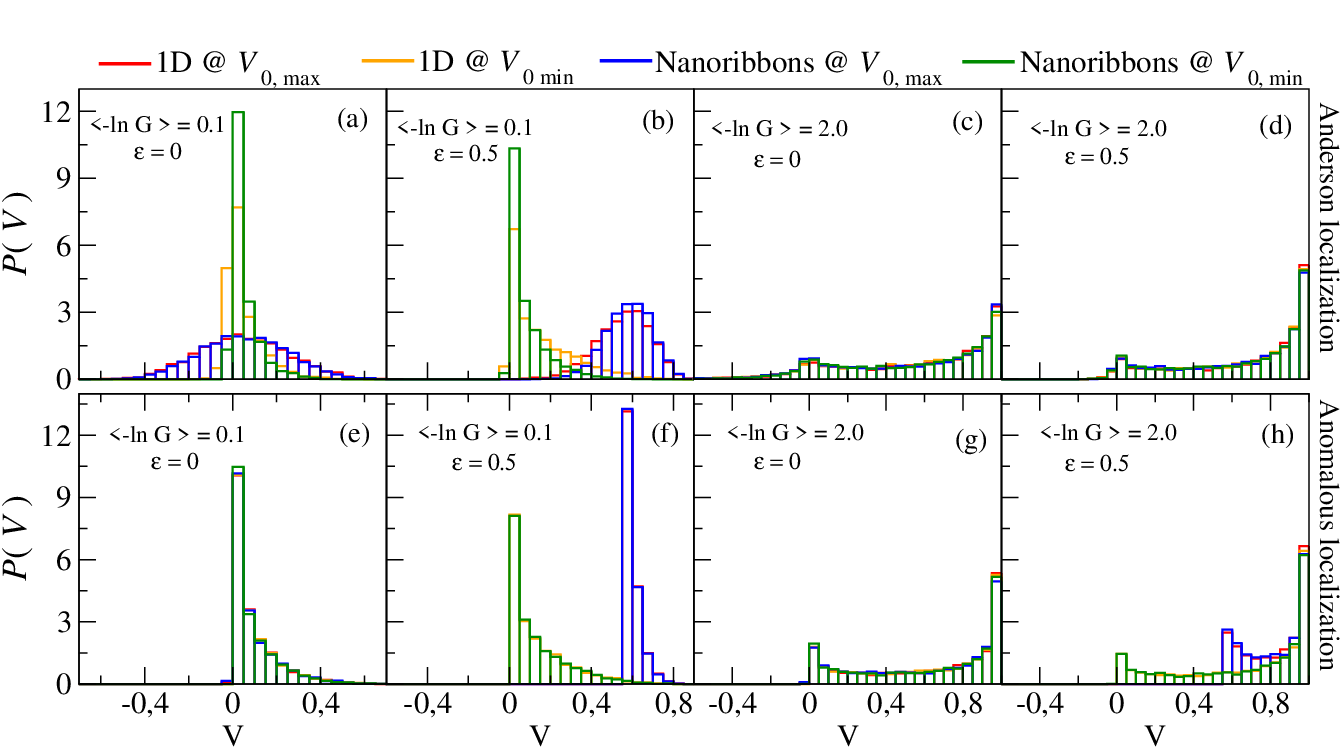}
\caption{Comparison of voltage distributions $P(V)$ obtained for 1D wires and nanoribbons, for two values of $\langle -\ln G \rangle$ ($0.1$ and $2.0$) and two values of the coupling $\epsilon$ ($0$ and $0.5$). Top panels show the results in presence of Anderson localization, and bottom panels in presence of anomalous localization. Each panel shows four cases: 1D wire for minimum voltage in absence of disorder, 1D wire for maximum voltage, nanoribbon for minimum voltage and nanoribbon for maximum voltage.}
\label{fig_6}
%\end{center}
\end{figure*}

We now consider the complete distribution of the voltage fluctuations. In the presence of Anderson localization, the voltage distributions for 1D wires and zigzag nanoribbons are shown in Fig.~\ref{fig_6}, upper panels, for two different degrees  of disorder characterized by $\langle -\ln G \rangle =0.1$ and 2.0, and two coupling values $\varepsilon =0$ and 0.5. The voltage distributions are calculated at $kL$ values where the maximum and minimum voltages, $V_{0, \mathrm{min}}$ and $V_{0, \mathrm{max}}$, as previously defined, are reached. As we can observe, the voltage distribution for zigzag nanoribbons and 1D wires are approximately the same for low disorder ($\langle -\ln G \rangle=0.1$), and they are indistinguishable for a higher amount of the  disorder ($\langle -\ln G \rangle=2$). 

In the lower panels of Fig.~\ref{fig_6}, the voltage distributions in the presence of anomalous localization are shown for armchair nanoribbons and 1D wires near the center of the band. As in the upper panels, the voltage distributions are obtained at two values of $kL$ at which the voltage reaches a maximum and minimum, $V_{0, \mathrm{min}}$ and $V_{0, \mathrm{max}}$, in pristine armchair nanoribbons. As we can see, the voltage distributions for 1D wires and armchair nanoribbons coincide.

Therefore, Fig.~\ref{fig_6} demonstrates that the type of localization, Anderson and  anomalous, plays a crucial role in  the statistics of the random fluctuations of the voltage. Furthermore, it shows that the obtained voltage statistics are not a consequence of the nanoribbons' lattices, but of the more general phenomenon of electron localization.

\section{Summary and conclusions}

We have studied the voltage fluctuations in disordered zigzag and armchair nanoribbons in a four-terminal configuration as a function of the distance between the voltage probes. The two terminals that act as voltage probes  are coupled to the ends of the nanoribbons  with   
coupling strength parameter $\varepsilon$. The presence of disorder in the  nanoribbon's lattices and multiple coherent  scattering processes of electrons lead to random voltage fluctuations. These voltage fluctuations are strongly dependent on the nanoribbon edges. The singular electronic properties of graphene lattices near the Fermi energy 
%at energies where a single channel contributes to the transport, 
allow us to study the effects of standard Anderson localization and anomalously localized electrons on the voltage fluctuations.

Away from the Fermi energy, the nanoribbons behave like ordinary metals, and we describe the voltage fluctuations using random matrix theory. In the ballistic regimen, the theoretical description is simplified, which allow us to provide analytical results for the average and variance of the voltage. We show that the average and  variance of the voltage oscillate with the distance between the voltage probes, and the voltage fluctuations can be so large that there is a nonzero probability of negative voltages. 
These results are confirmed by numerical simulations using the standard tight-binding model.   
We remark that negative resistances have been measured in single-wall carbon nanotubes in a four-configuration setup with a separation between voltage probes of the order of the mean free path \cite{Gao2005}.

Near the Fermi energy, electrons in disordered armchair nanoribbons are  weakly localized in relation to the exponential localization in the  standard Anderson phenomenon. In contrast, in zigzag nanoribbons, the electron wave functions are concentrated at the edge and are exponentially localized  in the presence of disorder. 
The power-law dependence of the average of the logarithmic conductance with the system length reveals the presence of anomalous localization. For anomalous localization in armchair nanoribbons $\langle \ln G \rangle \sim L^\alpha$ with $\alpha \approx 0.7$, whereas in zigzag nanoribbons, $\langle \ln G \rangle$ is a linear function of $L$, as in the standard Anderson localization phenomenon. 

When studying the effects  of these two types of localizations on the voltage, it was found that, in general, the voltage fluctuations are larger in zigzag nanoribbons than in armchair nanoribbons. Thus, the tails of the  broader voltage distributions in zigzag nanoribbons  are 
extended until negative voltage values. As a result, the probability of measuring negative voltages is finite in zigzag nanoribbons. Instead,   the support of the voltage probability distributions of armchair nanoribbons is 
restricted to positive voltage values. However, the coupling of the voltage probes reduces the voltage fluctuations in both types of nanoribbons. The most significant effects of the coupling of voltage probes are observed when the voltages are obtained at the value of the distance between probes at which the voltage reaches a maximum in its oscillatory behavior in pristine nanoribbons.

The voltage fluctuations studied here are the result of the general phenomenon of localization in quantum coherent transport. Thus, our results  
are not restricted  to  zigzag and armchair nanoribbons but are also applicable to disorder media where electrons are either Anderson localized or anomalously localized. As verification of this statement, we have shown that  the average voltage of zigzag and armchair nanoribbons match the average voltage in 1D disordered systems in the presence of  Anderson and nonstandard localizations. 

While the singular properties of nanoribbons are shown at low energies where a single channel contributes to the transport, it would be of interest to extend our analysis to the multichannel case and more general setup geometries, such as  voltage probes attached to  any point along the  nanoribbons. Additionally, it would be interesting to study the impact of different sources of disorder, such as vacancies or imperfections on the nanoribbon's edges, on voltage fluctuations.
Despite these  potential extensions of our study, the four-terminal setup we considered  is enough to demonstrate  essential coherent interference effects on the voltage.

\acknowledgements
V.A.G. acknowledges the financial support by MCIU (Spain) under the
Project No. PGC2018-094684-B-C22.

\end{document}